\documentclass[a4paper]{article}
\usepackage[dvips]{graphicx}
\usepackage{latexsym}
\usepackage{amssymb}
\usepackage{amsmath}
\usepackage{amsfonts}

\begin{document}

\title{Cryptanalysis of a novel cryptosystem based on chaotic oscillators and feedback inversion}
\author{
G. \'{A}lvarez Mara\~n\'on, L. Hern\'{a}ndez
Encinas\footnote{Corresponding author. Tel. (+34) 915 618 806
(Ext. 458), Fax: (+34) 914 117 651},\\
F. Montoya Vitini and J. Mu\~{n}oz Masqu\'e\\
\emph{Instituto F\'{\i}sica Aplicada}\\
\emph{Consejo Superior de Investigaciones Cient\'{\i}ficas}\\
\emph{C/ Serrano 144, 28006--Madrid, Spain}\\
\emph{Emails: \{gonzalo, luis, fausto, jaime\}@iec.csic.es}}

\date{}
\maketitle

\begin{abstract}
An analysis of a recently proposed cryptosystem based on chaotic
oscillators and feedback inversion is presented. It is shown how
the cryptosystem can be broken when Duffing's oscillator is
considered. Some implementation problems of the system are also
discussed.

\end{abstract}

\section{Introduction}

In recent years, a growing number of cryptosystems based on chaos
have been proposed \cite{Yang}, many of them fundamentally flawed
by a lack of robustness and security. In the Letter~\cite{SPG02},
the authors have proposed a symmetric cryptosystem based on
chaotic oscillators. More precisely, let $N\colon L_{\infty
e}\left( \mathbb{R}_{+}\right)  \rightarrow L_{\infty e}\left(
\mathbb{R}_{+}\right) $ be a non-linear time-varying system, where
$L_{\infty e}\left( \mathbb{R}_{+}\right)  $ stands for the
extended $L_{\infty}$-space on $\mathbb{R}_{+}$, and let $S_{1}$
and $S_{2}$ be two signal generators which generate the time
functions $t\mapsto w_{1}(t)$ and $t\mapsto w_{2}(t)$,
respectively. The encryption process is defined by adding the
signal generator $S_{2}$ to the output of the dynamic evolution of
$N$. Explicitly, to encrypt a message defined by a train of pulses
$t\mapsto p(t)$, of suitable width and of amplitude zero or one,
is necessary to compute the function $u(t)=p(t)+w_{1}(t)$,
$\forall t\geq0$. Then, $u(t)$ is applied to the system $N$ and
its output is added to $w_{2}(t)$. The ciphertext is then defined
to be $c(t)=(Nu)(t)+w_{2}(t)$, $\forall t\geq0$.

The decryption process consists of the two signal generators,
$S_{1}$ and $S_{2}$, and a feedback system $S(g,N)$, where $N$ is
the non-linear system used in the encryption protocol and $g$ is
the gain of the system. To decrypt a message $c(t)$, one subtracts
$w_{2}(t)$ from $c(t)$ and the result is the input to the system
$S(g,N)$. Its output, $\widetilde{u}(t)$, is a good, but noisy,
approximation of $u(t)$; also, the difference between $w_{1}(t)$
and $\widetilde{u}(t)$ gives $\widetilde{p}(t)$, which is a good,
but noisy, approximation of $p(t)$. Using a low-pass filter $L(s)$
and a quantizer $Q$, the original message is obtained. In order to
recover the original message exactly, $L(s)$ should be designed
carefully. Although the authors seem to base the security of their
cryptosystem on the chaotic behavior of the output of the non-linear
system $N$, no analysis of security is included.

In the present Letter we discuss the weaknesses of this system in
Section~\ref{sec:attacks} and we analyze its practical implementation in
Section~\ref{sec:practical}.

\section{Attacks to the cryptosystem}
\label{sec:attacks}

In this section we show how to break the cryptosystem proposed
in~\cite{SPG02} when Duffing's oscillator is used as the non-linear
time-varying system~\cite[\S 3.1]{SPG02}, which, in fact, is
the first example explained in detail.
The main problem with this cryptosystem lies on the
fact that the ciphertext is an analog signal, whose waveform
depends on the system parameter values and the plaintext signal.
Likewise, the detected signal before the quantizer depends on
these same parameters. The study of these signals provides the
necessary information to recover a good estimation of the system
parameter values and the correct plaintext.

We consider the first example in~\cite[\S 4.1]{SPG02}, for Duffing's
oscillator, represented by:
\begin{equation}\label{eq:duffing}
    N:\ddot x(t)+\delta\dot x(t)-\alpha x(t)+\beta
    x(t)^3=u(t),\;x(0)=0,\;\dot x(0)=0.
\end{equation}
In their example $w_1(t)=A\cos \omega t$ and $w_2(t)\equiv 0$. The
key of the system is made by the oscillator's parameters
$(\delta ,\alpha ,\beta)$, and the signal generator's parameters
$(A,\omega)$. Following the example given, we use a key formed by
the following set of parameters:
\begin{equation}\label{eq:realparameters}
    \alpha=10,\;\beta=100,\;\delta=1,\;A=1.5,\;\omega=3.76.
\end{equation}
Duffing's oscillator is used operating in the chaotic region.
This region is roughly characterized by the following values of
the parameters:
\begin{equation}\label{eq:chaotic}
3\leq\alpha\leq 15,\;40\leq\beta\leq 250,\;0.5\leq\delta\leq 1.7,\;
1\leq A\leq 2,\; 0.5\leq \omega \leq 7.
\end{equation}
The sensitivity to the parameter values is so low that the
original plaintext can be recovered from the ciphertext using a
receiver system with parameter values considerably different from
the ones used by the sender. As a consequence, it is very economic
to try different combinations of the parameters until a reasonable
approximation is reached. Although the parameter values can be
obtained with a very accurate precision, their knowledge is not
necessary to recover the plaintext.

We have found that the message can be decrypted even when $\beta$
has an error of $\pm 5\%$; $\delta$ has an error from $-30\%$ to
$+60\%$ and $\alpha$ has an absolute error of $\pm 2$ integers;
with respect to the original set of
parameters~\eqref{eq:realparameters}.

\begin{figure}[h]
  \center
  \includegraphics{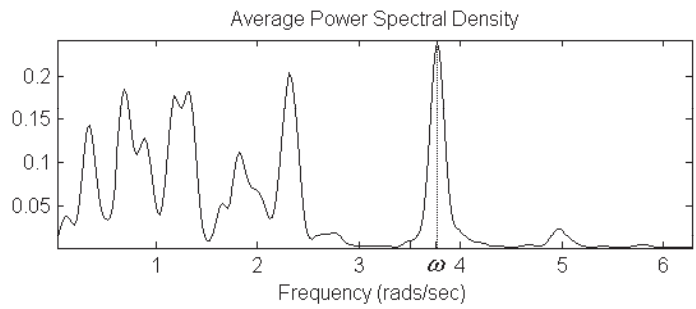}
  \caption{Power spectral analysis of the ciphertext
  signal. The highest peak corresponds to the frequency of $S_{1}$
  and lies at $\omega\approx 3.76$.}
  \label{fig:spectrum}
\end{figure}

In Fig.~\ref{fig:spectrum} we show the power spectral analysis
of the example ciphertext signal. As is observed, the frequency of
the forced oscillator is totally evident. The spectrum highest
peak appears at the $S_{1}$ signal generator frequency of
$\omega=3.76$. Thus, by simply examining the ciphertext, one of
the elements of the key ($\omega$) is obtained.

Next, the attacker uses a receiver for which $A=0$, and the rest of
the parameters takes values from the following sets:
\begin{align}
   \alpha&=\{5,9,13\}, \label{eq:alpha}\\
   \beta&=\{43,47,51,56,62,68,75,82,91,100,110,\nonumber\\
        &\quad \quad 120,130,145,160,180,200,220,240\}, \label{eq:beta}\\
   \delta&=\{0.7, 1.3\}.\label{eq:delta}
\end{align}
This makes a total of 114 possible combinations, which should be
tried one by one. To check whether the choice of the parameters is
good, we look at the output of the low-pass filter $L(s)$, which
we call $\hat p(t)$. When the parameter values are slightly
different from the right ones, then $\hat p(t)$ will look like a
square signal summed with a pure sine. The frequency of this sine
corresponds to the value of $\omega$ previously calculated from
the spectrum of the ciphertext. The amplitude of this sine
corresponds exactly to the value of $A$ used by the sender.

\begin{figure}[h]
  \hspace{-1.5cm}
  \includegraphics{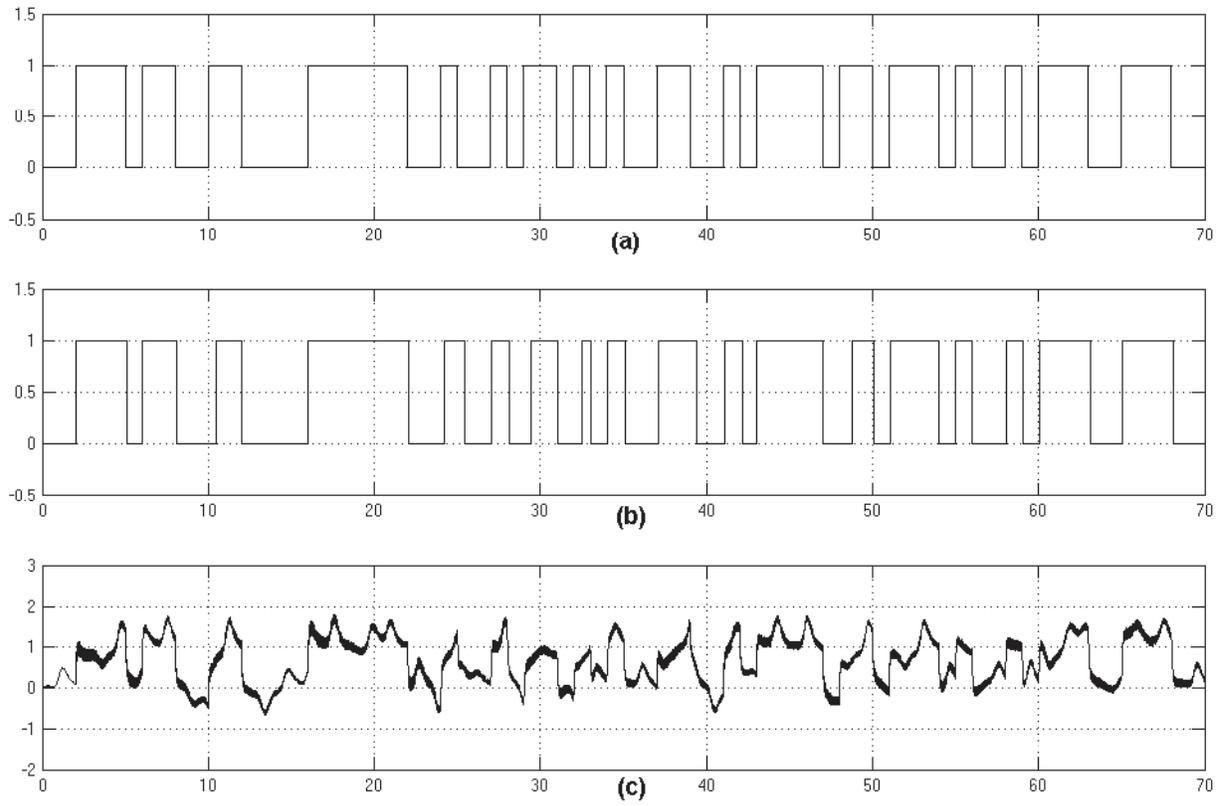}
  \caption{Plaintext recovery with inexact receiver
  parameter values. Time histories of: (a) plaintext; (b) recovered
  plaintext; (c) $\hat p(t)$.}
  \label{fig:histories}
\end{figure}

Next, the value of $A$ just computed is used to regenerate the
plaintext. Due to the low sensitivity to the parameter values,
although the exact values are unknown, the deciphered plaintext
signal will be equal (or very close) to the original one. In
Fig.~\ref{fig:histories} the recovered plaintext is depicted for
the following parameter values:
\begin{equation}\label{eq:guessedparameters}
    \alpha=9,\;\beta=100,\;\delta=0.7,\;A=1.4,\;\omega=3.76.
\end{equation}
The first three values are taken from the
equations~\eqref{eq:alpha}--\eqref{eq:delta}. Although the parameter
errors are $10\%$, $0\%$, $30\%$, $6.66\%$, and $0\%$, respectively,
the plaintext is correctly recovered. These values could be further
refined by varying them in an effort to approximate $\hat p(t)$ to
a square wave.

\section{Difficulties of practical implementation}
\label{sec:practical}

In this section we discuss the difficulties that this cryptosystem
will face if it is practically implemented.

\subsection{Analog transmission}

The proposed cryptosystem seems to present serious problems in
a real transmission, because the
recovered signal at the receiving end of the transmission path
will be very difficult to decrypt.

Apparently, the authors have only implemented a software
simulation of the complete encryption/decryption system, feeding
the ciphertext (the output generated by the encryption system)
directly as the input to the decryption system. The generators
$S_1$ and $S_2$, part of the system key, look to be connected
simultaneously and locally to both the encryption and decryption
systems.

In real world applications, however, things happen in a very
different way. Ideal transmission lines introduce an unknown
amount of attenuation and delay in the transmitted signal.
Furthermore, real transmission lines introduce distortion and
noise too. Moreover, wireless communication systems exhibit
time-variable attenuation and delay.

Thus, the input signal at the receiver end $c'(t)$ and the
transmitted signal $c(t)$ will differ. In the most favorable case,
if we assume that we are using an ideal line, the received signal
will be $c'(t)= k c(t+\tau)$, were $k$ and $\tau$ are the
attenuation and delay of the line.

\subsubsection{Synchronization}

As the authors point out, most continuous chaotic cryptosystems
described until now are based on the synchronization of two
chaotic systems. The claimed novelty of the present cryptosystem
relies on the lack of synchronization between encryption and
decryption; but this is an erroneous claim, because in the software
simulations the authors have used a hidden synchronization
mechanism consisting of the local and simultaneous connection of
generators $S_1$ and $S_2$ to both the encryption and decryption
systems.

In real world applications, given that transmission lines have
limited bandwidth, when transmitting to a remote system, signal
delay will take place. The delay amount may vary for different
frequency components of the signal, depending on the line
impulsive response. Thus, the observed waveforms at sender and
receiver ends may differ and it will be very difficult to estimate
the right moment to start the receiving generators.

Some measures should be taken to assure that both ends are using
signal generators $S_1$ and $S_2$ with exactly the same phase in
respect to the ciphertext. However, no mesure is considered by the
authors. Hence the receiver end's generators will never generate
the adequate signal.

\subsubsection{Attenuation}

Another factor to be considered is the line attenuation. No
continuous transmission or storage system (cable, optical,
magnetic or wireless) grants that the received or reproduced
signal amplitude preserves the original amplitude. If the signal
is transmitted over a switched network, the attenuation will change
each time that a new connection is made. If the signal is
transmitted over a wireless channel, the attenuation will vary
depending on changing atmospheric conditions, changing
reflections, and changing multipath.

When transmitting a signal of known constant amplitude  (e.g.
square pulses or frequency modulated sinusoids) it is possible to
equalize the received signal, restoring the correct amplitude
level. But in the present case, as the signal is chaotic, its
amplitude is varying in an unpredictable fashion, so it is
impossible any level restoring.

Therefore, it will be impossible to subtract exactly $w_2$ at the
receiver end. Hence, the signal $r$ at the input of the decryption
system feedback loop will be $r(t)= c'(t)-w_2(t)$, i.e.:
\begin{equation}\label{eq:r}
r(t)=k(Nu)(t+\epsilon)+kw_2(t+\epsilon)-w_2(t),
\end{equation}
where $\epsilon$ is the time inaccuracy in the determination of
the right moment to start the receiving generators.
Hence, the decrypted signal will be:
\begin{equation}\label{eq:y}
y(t)\approx(N^{-1}r)(t)=(N^{-1}(k(Nu)(t+\epsilon)+kw_2(t+\epsilon)
-w_2(t)))(t).
\end{equation}
As $N$ is a nonlinear chaotic map and due to the \emph{sensitive
dependence on the initial conditions} that characterize
chaos~\cite[p. 119]{Devaney92}, the decrypted signal
$y(t)$ will never match the plaintext $p(t)$.

\begin{figure}[h]
\begin{center}
  \includegraphics{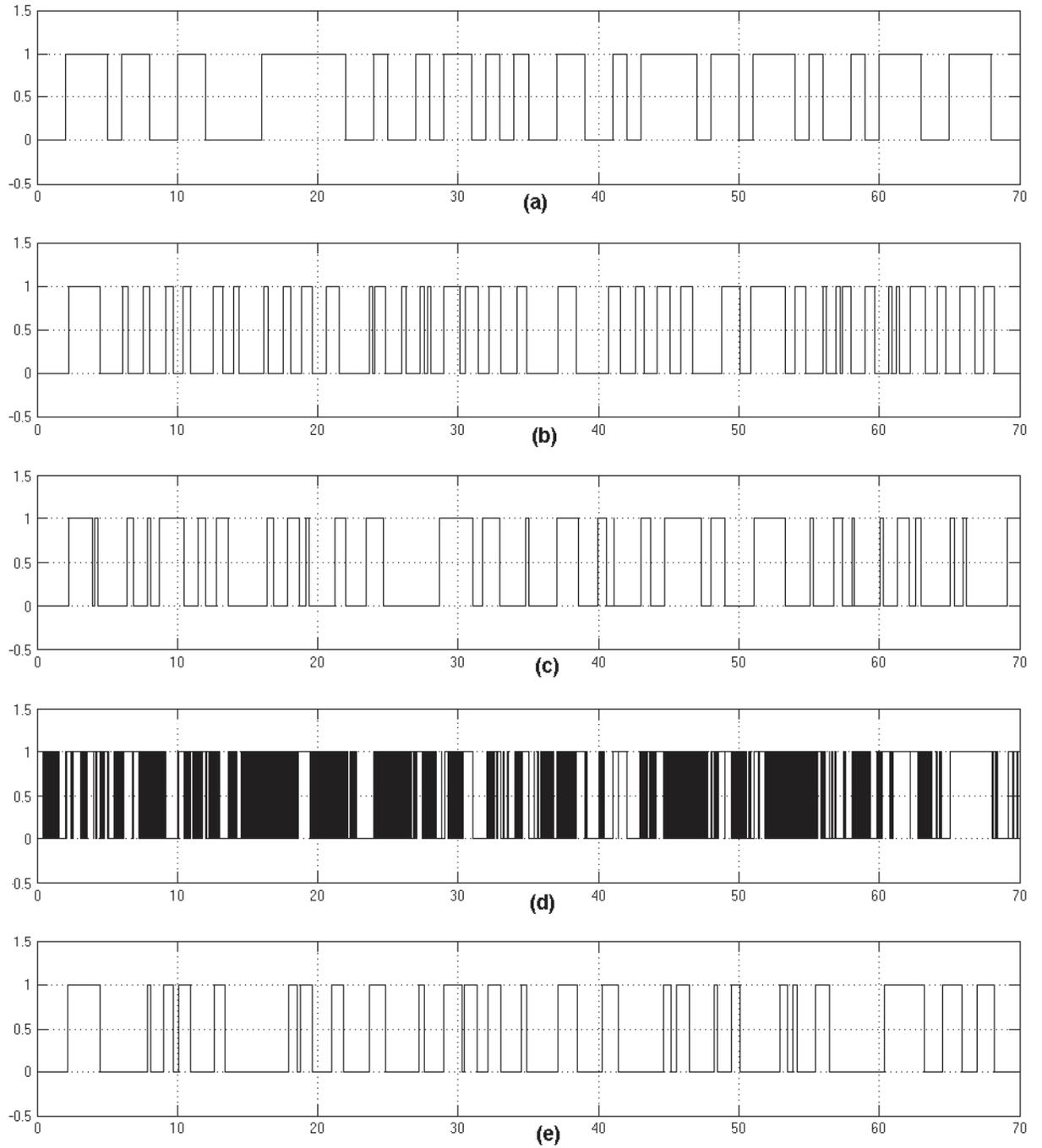}
  \caption{Effects of a real communication channel: (a) plaintext;
  (b) recovered plaintext with channel bandwidth restricted to
  $\omega=6.28$~rad/seg; (c) recovered plaintext with channel attenuation of
  $3$~dB; (d) recovered plaintext with channel noise of $-40$~dB;
  (e) recovered plaintext with channel bandwidth restricted to
  $\omega=9.42$~rad/seg, attenuation of $0.5$~dB and noise of $-50$~dB. The
  parameter values at the sender and receiver ends match exactly.
}
  \label{fig:channel}
\end{center}
\end{figure}

The recovered plaintext errors induced by the use of a real
communication channel with restricted bandwidth, attenuation
and/or noise are illustrated in Fig.~\ref{fig:channel}.

Moreover, the authors seem to base the security of their
cryptosystem on the chaotic behavior of $N$, although no evidence
of that is shown. In any case, the chaotic profile of the output
$x(t)$ in Duffing's oscillator~\eqref{eq:duffing} is not always
guaranteed for every input $u(t)=p(t)+w_1(t)$, even in the chaotic
range~\eqref{eq:chaotic}, and the sensitive dependence on the
initial conditions is diminished as they are kept to be fixed,
$x(0)=0$, $\dot x(o)=0$, in Duffing's equation~\eqref{eq:duffing}.

\subsection{Digital transmission}

If a discretization of the ciphertext is sent instead of the
dynamic evolution of the system $N$, then there are two options.

In the first one, the ciphertext is discretized only at the nodes
$i=0,\ldots,n$, where $n$ is the number of pulses of $p(t)$. Then,
the ciphertext sent by the sender must be the $3$-uples
$(x(t_{i}), \dot{x} (t_{i}),\ddot{x}(t_{i}))$, $i=0,\ldots,n$,
since the receiver needs to know these values---and not only the
$x(t_{i})$---in order to be able to decrypt the message, as the
usual methods of discretization do not allow to obtain the values
of the derivative at the nodes $t_{i}$ in terms of the values of
the function at such nodes. For example, if one uses the
Runge-Kutta method (see \cite[\S163]{Zwillinger89}) to solve
$\ddot{x}=f\left( t,x,\dot{x}\right)$, then the values of the
first derivative are given by
\begin{align}
\dot{x}(t_{0}+h) & =\dot{x}(t_{0})+\frac{1}{6}\left( k_{1}+2k_{2}
+2k_{3}+k_{4}\right) ,\\
k_{1} & =hf\left( t_{0},x\left( t_{0}\right) ,\dot{x}\left(
t_{0}\right)
\right) ,\\
k_{2} & =hf\left( t_{0}+\frac{1}{2}h,x(t_{0})+\frac{1}{2}h\dot{x}
(t_{0})+\frac{1}{8}hk_{1},\dot{x}(t_{0})+\frac{1}{2}k_{1}\right) ,\\
k_{3} & =hf\left( t_{0}+\frac{1}{2}h,x(t_{0})+\frac{1}{2}h\dot{x}
(t_{0})+\frac{1}{8}hk_{2},\dot{x}(t_{0})+\frac{1}{2}k_{2}\right) ,\\
k_{4} & =hf\left(
t_{0}+h,x(t_{0})+h\dot{x}(t_{0})+\frac{1}{2}hk_{3}, \dot
{x}(t_{0})+k_{3}\right) .
\end{align}
We remark on the fact that the values for the second derivative
should be computed from the formula
\begin{equation}
\ddot{x}(t_{i})=f\left( t_{i},x(t_{i}),\dot{x}(t_{i})\right) ,
\end{equation}
where $f$ is the function defining the dynamic system $N$. This
fact implies that the transmission has a high factor expansion as
every pulse of the original message is transmitted by means of a
$3$-uple of real numbers with a consistent number of decimals.

The second option consists in computing a much more long list of
values $x(s_{i})$, $i=0,\ldots,m$, with $m\gg n$. In this case,
the values for the first derivative can be obtained from the
formulas above;\ but the second derivative should also be included
in the transmission. Hence, in this case the ciphertext is
$(x(s_{i}),\ddot{x}(s_{i}))$, $i=0,\ldots,m$. What is gained in
not sending the first derivative is lost by the greater number of
entries of the list.

In any case, the values of the first and second derivatives cannot
be computed by the usual approximate formulas
\begin{align}
\dot{x}(t_{i}) & \approx\frac{x(t_{i+1})-x(t_{i})}{h},\\
\ddot{x}(t_{i}) &
\approx\frac{x(t_{i+2})-2x(t_{i+1})+x(t_{i})}{h^{2}},
\end{align}
as they produce considerable errors in the decryption process due
to the nonlinear terms in $N$.

\section{Conclusion}
As a consequence of the previous analysis, the cryptosystem studied
cannot work in practice because it is not using any synchronization
mechanism and because it is not robust to real channel conditions.
On the other hand, the cryptosystem is rather weak, since it can
be broken by using a set of $114$ parameter values only. The total
lack of security, along with the lack of robustness, discourages
the use of this algorithm for secure applications.

\vspace{0.5cm}
\noindent {\bf Acknowledgements} This work is supported by
\textit{Ministerio de Ciencia y Tecnolog\'{\i}a} of Spain,
research grant TIC2001-0586.

\clearpage
\pagestyle{empty}

\section*{Figure captions}

\clearpage

\clearpage

\end{document}